\documentclass[conference]{IEEEtran}
\IEEEoverridecommandlockouts
\usepackage{cite}
\usepackage{amsmath,amssymb,amsfonts}
\usepackage{algorithmic}
\usepackage{graphicx}
\usepackage{textcomp}
\usepackage{xcolor}
\def\BibTeX{{\rm B\kern-.05em{\sc i\kern-.025em b}\kern-.08em
    T\kern-.1667em\lower.7ex\hbox{E}\kern-.125emX}}
    
\usepackage{xcolor}

\begin{document}

\title{Extending SOUP to ML Models When Designing Certified Medical Systems}

\author{\IEEEauthorblockN{Vlad Stirbu}
\IEEEauthorblockA{
\textit{CompliancePal}\\
Tampere, Finland \\
vlad.stirbu@compliancepal.eu}
\and
\IEEEauthorblockN{Tuomas Granlund}
\IEEEauthorblockA{
\textit{Solita}\\
Tampere, Finland \\
tuomas.granlund@solita.fi}
\and
\IEEEauthorblockN{Jere Helén}
\IEEEauthorblockA{
\textit{University of Helsinki}\\
Helsinki, Finland \\
jere.helen@helsinki.fi}
\and
\IEEEauthorblockN{ Tommi Mikkonen}
\IEEEauthorblockA{
\textit{University of Helsinki}\\
Helsinki, Finland \\
tommi.mikkonen@helsinki.fi}
}

\maketitle

\begin{abstract}
Software of Unknown Provenance, SOUP, refers to a software component that is already developed and widely available from a 3rd party, and that has not been developed, to be integrated into a medical device. From regulatory perspective, SOUP software requires special considerations, as the developers' obligations related to design and implementation are not applied to it. In this paper, we consider the implications of extending the concept of SOUP to machine learning (ML) models. As the contribution, we propose practical means to manage the added complexity of 3rd party ML models in regulated development.
\end{abstract}


\section{Introduction}

Modern software intensive products are complex systems that often incorporate software components developed by third parties. Across the board, we can see that embedded systems rely on open source operating systems, mobile applications rely heavily on the operating systems and the middleware developed by the device manufacturers, cloud applications depend on the cloud infrastructure and services maintained by the cloud provider, while web applications rely on the web browsers and their APIs. Moving into the application stack, there are countless open source libraries and frameworks that are used to develop applications. The high level of software reuse has enabled organizations to efficiently develop sophisticated products at a fast pace. While differentiating features are implemented internally, most of the other components of a software product are developed by a third party.

Due to safety considerations, the medical industry has a very conservative stance on how the software is developed and used. Applications that incorporate artificial intelligence (AI) and machine learning (ML) technologies are becoming popular due to their ability to build complex prediction systems. The abilities enabled by the ML models complicates further the software development process. They bring additional risks such as increased complexity, lack of transparency, danger of inappropriate use, or weak governance that have to be handled appropriately from a safety perspective. 

In this paper we explore the implications of using ML models developed by third parties in software intensive medical devices, from a regulatory perspective. First, we look at the new challenges introduced by the ML technologies in medical applications. Then, we propose practical means to manage the added complexity of 3rd party ML models in regulated development. Towards the end of the paper, we finally draw some conclusions.


\section{Background}

\subsection{Medical regulatory framework}

The regulations covering medical device software fall into two broad categories: information handling and safety. The regulations related to information handling are Health Insurance Portability and Accountability Act (HIPAA) \cite{HIPAA} in United States and General Data Protection Regulation (GDPR) \cite{GDPR} in the European Union. HIPAA is a medical sector regulation that defines what constitutes \textit{protected health information}, its use and disclosure when several health providers are involved in the care process. GDPR is a generic data privacy framework that defines how personal information is collected and used. To comply with HIPAA and GDPR regulatory frameworks, a service provider that implements part of the functionality using software must establish procedures for handling the relevant information. These procedures typically get materialized into technical requirements, which have to be implemented in software, or standard operating procedures, which have to be followed by the staff that interacts with the protected information. The regulations extend to business associates that process or handle protected information, which have to comply themselves.

The safety of medical devices or services is regulated in Unites States by the Food and Drug Administration (FDA) and Medical Device Regulation (MDR) \cite{MDR} in European Union. International Organization for Standardization (ISO) and International Electrotechnical Commission (IEC) have developed international standards that contain guidance on the processes and requirements that must be followed when developing software so that relevant regulatory authorities  accept medical products in the respective markets. 

For example, ISO 13485 \cite{ISO-13485} specifies the requirements for a quality management system that allows an organization to demonstrate its ability to provide medical devices and related services that consistently meet customer and applicable regulatory requirements. Further, ISO 14971 \cite{ISO-14971} specifies the processes that a manufacturer must follow to identify the hazards associated with medical devices, to estimate and evaluate the associated risks, to control these risks, and to monitor the effectiveness of these controls. Finally, the life cycle requirements that must be followed by an organization where software is embedded or is an integral part of the final medical device are defined in IEC 62304 \cite{IEC-62304}. The requirements are envisioned as a set of processes, activities and tasks that establish a common framework for medical device software life cycle processes.

\subsection{ML development lifecycle and challenges}

\begin{figure}
    \centering
    \includegraphics[width=0.45\textwidth]{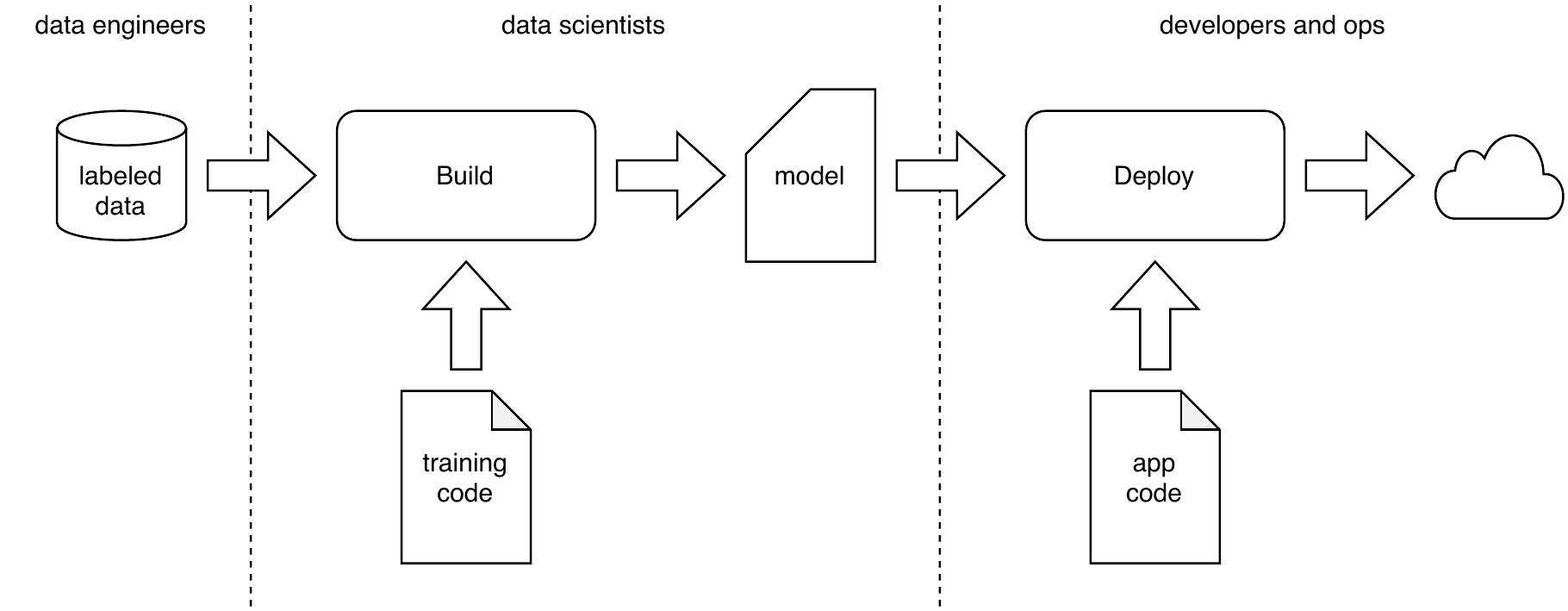}
    \caption{Functional silos barriers when developing ML applications}
    \label{fig:ml-silos}
\end{figure}

Building on the success of continuous software development approaches \cite{fitzgerald2014continuous,fitzgerald2017continuous}, in particular DevOps \cite{lwakatare2015dimensions}, it has become desirable to deploy machine learning (ML) components in real time, too. To this end, MLOps refers advocating automation and monitoring at all steps of ML system development and deployment, including integration, testing, releasing, deployment and infrastructure management.


To understand the challenges related to MLOps, let us first explain the steps necessary to train and deploy ML modules \cite{SWQD21}. As the starting point, data must be available for training. There are various somewhat established ways of dividing the data to training, testing, and cross-validation sets. Then, an ML model has to be selected, together with its hyperparameters. 
Next, the model is trained with the training data. During the training phase, the system is iteratively adjusted so that the output has a good match with the ``right answers'' in the training material. This trained model can also be validated with different data. If this validation is successful -- with any criteria we decide to use -- the model is ready for deployment, similarly to any other component. Once deployed, ML related features need monitoring, like any other feature. However, monitoring in the context of ML must take into account inherent ML related features, such as biases and drift that may emerge over time. In addition, there are techniques that allow improving the model on the fly, while it is being used. Hence, the monitoring system must take these needs into account.



Based on the above, continuous deployment of ML features is often a complex procedure that involves changes in the following areas: the application code, the model used for prediction, and the data used to develop the model. Often, these areas are handled separately by software developers, data scientists and data engineers that rely on different skill sets and tool chains. For example, data engineers are focused on making the data more accessible, data scientist perform experiments for improving the data model, and the developers are worried about integrating the various technologies and releasing them to production (Figure \ref{fig:ml-silos}). The lack of harmonized processes across these domains leads to delays and frictions, such as models never reaching production or deployments that are difficult to update or debug. Due to this variability, machine learning applications are more complex than traditional applications. These characteristics makes them harder to test, explain or improve.

\subsection{General purpose MLOps pipelines}

\begin{figure}
    \centering
    \includegraphics[width=0.45\textwidth]{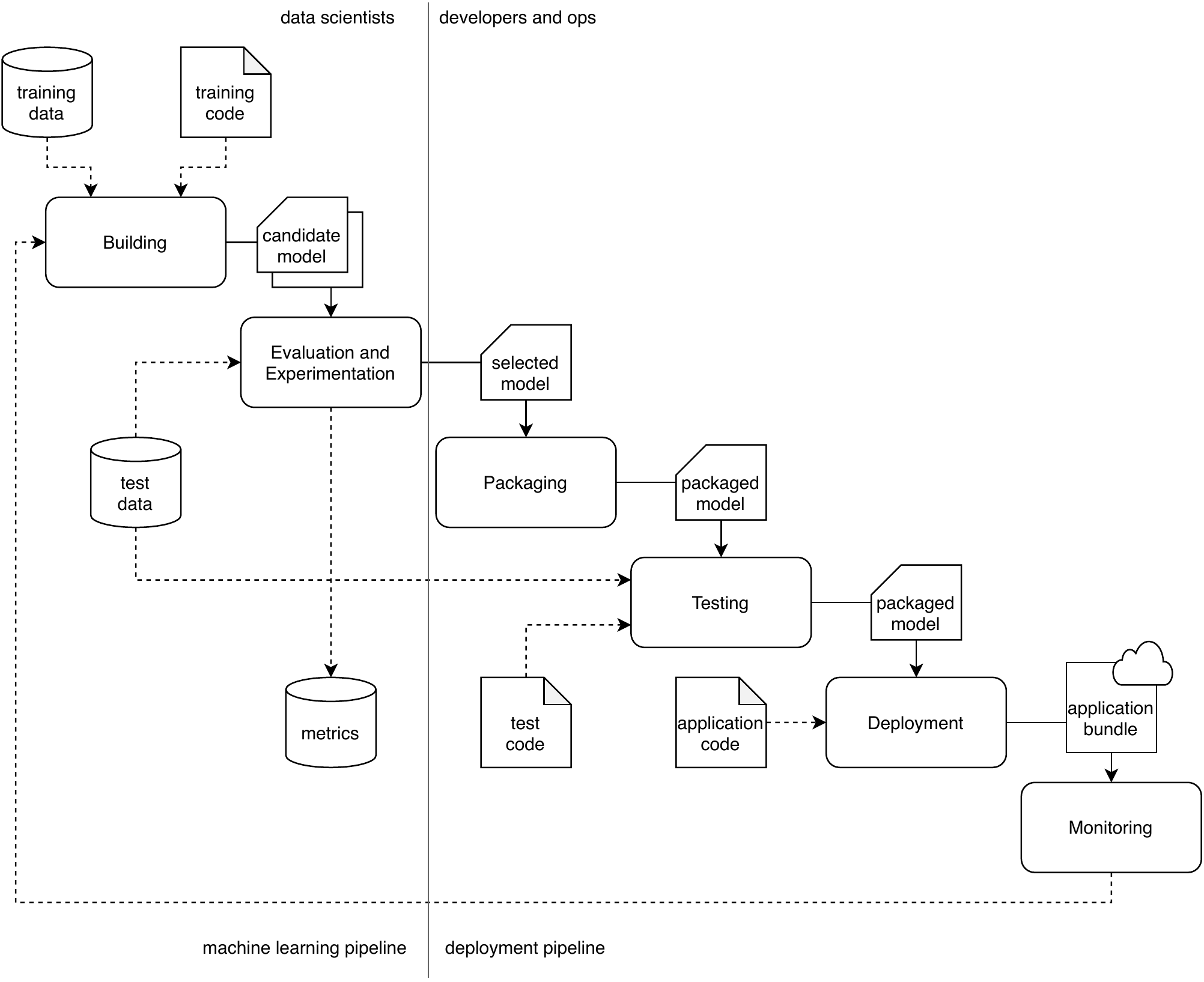}
    \caption{CD4ML pipelines and artifacts}
    \label{fig:cd4ml}
\end{figure}

Continuous Delivery for Machine Learning (CD4ML) \cite{cd4ml} is an approach formalized by ThoughtWorks for automating in an end-to-end fashion the lifecycle of machine learning applications. In CD4ML a cross-functional team produces machine learning applications based on code, data, and models in small and safe increments that can be reproduced and reliably released at any time, in short adaptation cycles. The approach contains three distinct steps: identify and prepare the data for training, experimenting with different models to find the best performing candidate, and deploying and using the selected model in production. This has been illustrated in Figure \ref{fig:cd4ml}.

The first step has the goal of making the data \textit{discoverable and accessible}. It consists in collecting relevant data from different internal and external sources, transforming and exposing it in a format that is used by the data scientists to train the model. The data pipeline codifies the directed acyclic graph that contains the sources, the destinations and the transformations performed on the data. As the source data used in this step can be very large, it is not practical to check it in version control. Instead, metadata that conveys the location, ranges and other parameters that determine that shape of the data source used for training. Over time the data can evolve over two axes: data schema or sampling frequency. Storing the pipeline, the source data metadata and the code that performs the transformations is an effective data provenance mechanism.

The next step is to train model candidates based on the data collected in the previous step. The input data is split into training and validation data. The training data is used to evaluate combinations of algorithms that produces a model. The model is evaluated against the validation set to assess its quality. This process is codified as the \textit{machine learning} pipeline. During development the pipeline can change frequently and it is difficult to reproduce the process outside the local environment without the assistance of specialized tools 
that provide git-like functionality that keeps track of data and code used in experiments, allowing execution on other environments. As most experiments do not yield good results, it is critical to preserve all data, metadata, metrics and the code that captures how the experiment was conducted. This record supports the decision process for promoting a particular model to production.

The last step is to deliver the model into the production environment using a \textit{deployment pipeline}. The process consists in testing the model selected for production, packaging in the format suitable for production, followed by deployment. The deployment can follow one of the following patterns: include in the application code as normal dependency, run it as standalone service, or deploy at runtime as data. A special case of deployment is \textit{online learning}, where the models constantly learn in production. Versioning the model as static artifact won’t yield the same results as they are fed different data. Once in production, continuous monitoring ensures that the model behaves as expected, and anomalies are detected and handled properly. The feedback loop allows the model to be improved over time using observations from production environment.

\subsection{Risk management for ML applications}

Algorithms are an essential ingredient of machine learning. The risks inherent to algorithm design propagate to medical machine learning applications due to their increased complexity, lack of transparency, inappropriate use, or weak governance. According to Krishna et al. \cite{ml-risk-framework}, algorithmic risk can be split into three categories: input data, algorithm design and output decisions. Flaws in input data such as biases in the data used for training, the quality of the data can lead to mismatches between the data used for training and the data used during normal use. Output decisions flaws relate to incorrect interpretation or use of the output. Algorithm design flaws can be expanded in human biases – cognitive biases of model developers and users can lead to flawed output, technical flaws – lack of technical rigour or conceptual soundness during development, training, testing and validation, usage flaws – incorrect implementation or integration in operations can led to inappropriate decision-making, or security flaws – threat actors can manipulate the inputs to produce deliberate flawed outputs.

As new machine learning techniques are so opaque, it is practically hard to understand how they operate. Industry groups \cite{aiethics} and academia recognised the challenges associated with machine learning trustworthiness. Their proposals to improve trustworthiness include metrics like interpretability - degree to which a human can understand the cause of a decision \cite{doshivelez2017rigorous}, and explainability -- the degree to which a human understands the behaviour of a system \cite{explainableai}, or frameworks that allow to explain machine learning models as black \cite{ai-blackbox1}, \cite{interpretable-ml}, or white boxes \cite{ml-whitebox}. 

\section{Discussion}

\subsection{SOUP handling under IEC 62304}

\begin{figure}
    \centering
    \includegraphics[width=0.40\textwidth]{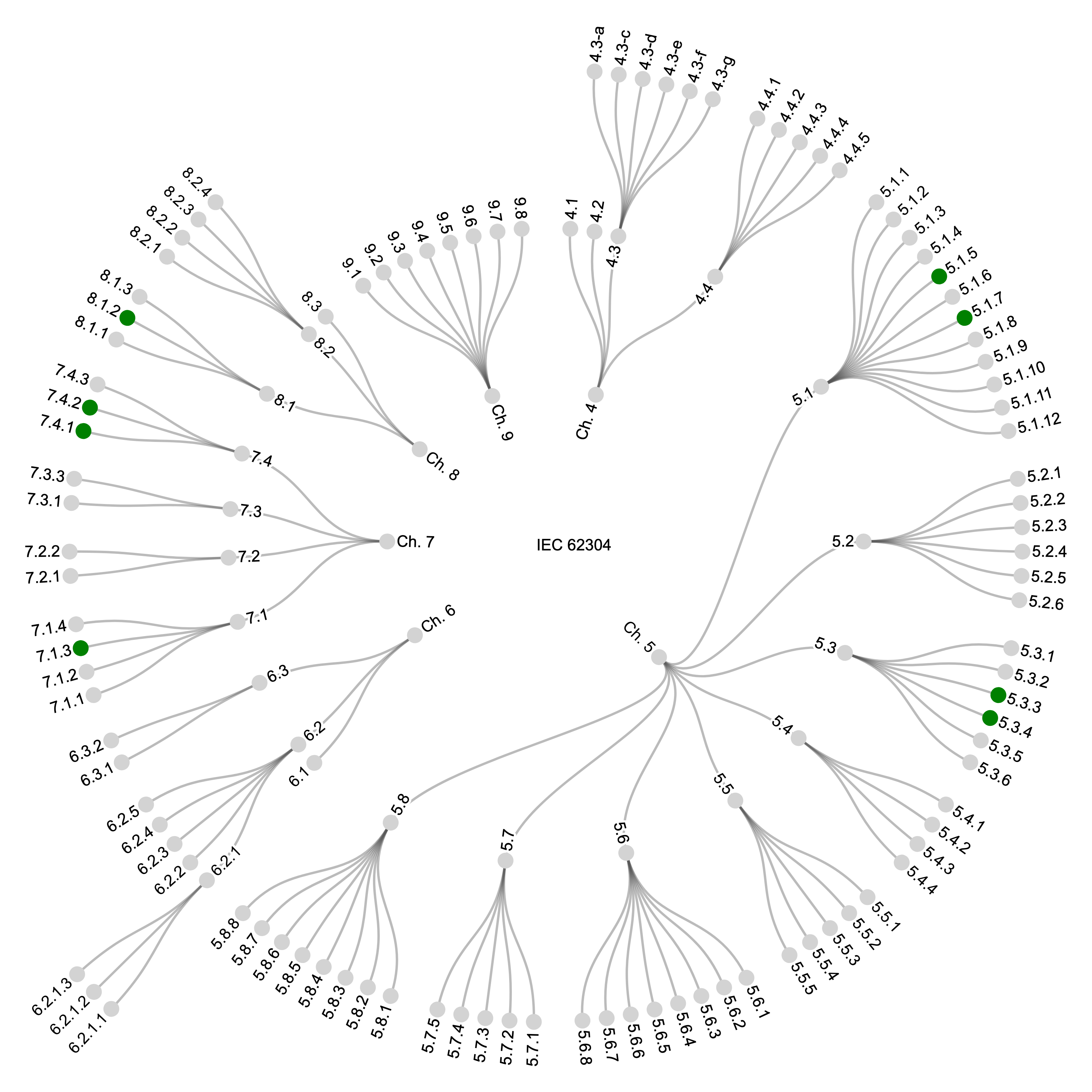}
    \caption{SOUP related clauses in IEC 62304}
    \label{fig:soup-clauses}
\end{figure}

Medical device manufacturers are responsible for the safety of their products, and the implementation of the appropriate procedures that ensure the safety. Modern software intensive devices and services incorporate often components developed by third parties that have not followed the software development lifecycle requirements described by IEC 62304. From a regulatory perspective, the third party software is considered software of unknown provenance (SOUP). To be able to use the SOUP components, manufacturers need to plan the integration, testing (clause 5.1.5), and risk management activities (clause 5.1.7), and specify the functional, performance and hardware requirements (clauses 5.3.3 and 5.3.4). Further, they need to continuously evaluate the anomalies published by the maintainer to asses if they impact the product safety (clause 7.1.3), and perform risk management activities for the changes caused by the SOUP components (clauses 7.4.1 and 7.4.2). Finally, the manufacturer needs to maintain a registry with SOUP components used in the product.

Although the ratio of SOUP specific clauses is small (see Fig. \ref{fig:soup-clauses}), the resources that must be allocated by the manufacturer could be considerable. While the SOUP components can speed the development considerably, they increase the complexity of the overall software system and extend the surface on which attack vectors could manifest. Understanding the intended purpose for which the SOUP component was developed and how well it matches the usage context within the medical products is just a first step. Quite often, identifying the safety risks and the controls that mitigate them is complemented by other activities that asses the engineering practices maturity used during development, identify threats in the area of cybersecurity, or legal restrictions due to the SOUP component's license.

\subsection{Extending the SOUP to ML models}

ML models can be integrated into a product using one of the following approaches: (i) as a library dependency, (ii) as a service, or (iii) as data loaded at runtime. Technically, all these approaches qualify the third party developed ML model as a SOUP component. The manufactures should include the component into the software decomposition as an item, and perform the appropriate software integration and risk management activities specified in the plans. Furthermore, the manufacturer must ensure that besides the packaged ML model, the accompanying documentation provided by the model developer includes the appropriate explanations and to a level of detail that allows the effective mitigation of the risks associated with ML applications described earlier. This documentation should describe at least the input data, the algorithm design and the output decisions. If available, a validation data set should be used in integration tests to ensure that the ML model is integrated correctly. Additionally, the third party models should be evaluated from a cybersecurity perspective to ensure that new threats are not silently introduced into the medical system \cite{ai-cyber-security}.

From a continuous software engineering perspective, medical ML applications require a more complex machinery for handling the software development lifecycle. If we look at the CD4ML, we see that data used to develop the algorithms, but also the ML experiments themselves have to be version controlled, not only the code. These artifacts have not been so far in the scope of the medical regulations. Conventions must be established in order to facilitate effective handling and use of the artifacts created at each step of the pipelines during regulatory activities.

Finally, to add complexity to continuous delivery pipeline, ML applications can introduce a need for additional functions in operational use, such as monitoring the operations and detecting possible bias that can emerge once the system has been deployed. Obviously, such monitoring mechanisms, together with the infrastructure used for reporting are subject to regulatory actions. In addition, ensuring that monitoring is compatible with the ML components that have been deployed must be ensured. Validating this may require access to training data sets as well as the models themselves. 

\section{Conclusions}

Large scale software reuse of third party components speeds up the product development considerably. From a regulatory perspective, the SOUP components must be handled according to special procedures to ensure that they do not affect negatively the product's safety. Machine learning models enable complex predictions systems with benefits in medical applications, but at the cost of increased complexity and added specific risks. In this paper we advocate for extending the regulatory SOUP procedures and activities to ML models. Additionally, we propose practical guidelines for evaluating the engineering practices used for developing the models, as well as data that must be provided downstream, so that medical manufacturers can integrate them and perform their safety evaluation effectively.


\textbf{Acknowledgments}. The authors would like to thank Business Finland and the members of AHMED (Agile and Holistic MEdical software Development) consortium for their contribution in preparing this paper.

\bibliographystyle{./bibliography/IEEEtran}
\bibliography{./bibliography/paper}

\end{document}